

\magnification=\magstep 1
\vsize=9.5 true in
\baselineskip=14pt
\font\bhead=cmbx10 scaled 1440
\def \d {{\rm d}}

\pageno=0
\footline={\ifnum\pageno>0\hss\rm\folio\hss\else\hfill\fi}

\noindent July 1997 
\vskip 1.0cm
\centerline{\bhead Gravitational waves with distinct wavefronts}

\bigskip\centerline{by}\bigskip
\centerline{{\bf G. A. Alekseev}\footnote{$^1$}
{e-mail: {\tt G.A.Alekseev@mi.ras.su}}}
\centerline{Steklov Mathematical Institute}
\centerline{Gubkina 8, Moscow 117966, GSP-1, Moscow, Russia}

\bigskip\centerline{and}\bigskip
\centerline{{\bf J. B. Griffiths}\footnote{$^2$}
{e-mail: {\tt J.B.Griffiths@Lboro.ac.uk}}}
\centerline{Department of Mathematical Sciences}
\centerline{Loughborough University, Loughborough, Leics. LE11 3TU. U.K.}
\vskip 1.0cm

\noindent{\bf Abstract}
\medskip
Exact solutions of Einstein's vacuum equations are considered which describe
gravitational waves with distinct wavefronts. A family of such solutions
presented recently in which the wavefronts have various geometries and which
propagate into a number of physically significant backgrounds is here related
to an integral representation which is a generalisation of the Rosen pulse
solution for cylindrical waves. A nondiagonal solution is also constructed
which is a generalisation of the Rosen pulse, being a cylindrical pulse wave
with two states of polarization propagating into a Minkowski background. The
solution is given in a complete and explicit form. A further generalisation
to include electromagnetic waves with a distinct wavefront of the same type
is also discussed.

\vfil

\hfill  PACS class - 04.30.Nk, 04.20.Jb

\vfil\eject
\noindent{\bf 1. Introduction}
\medskip
In many physically realistic situations concerning gravitational wave
emission and propagation, it seems natural to expect the occurrence of waves
with distinct wavefronts. By contrast, most exact solutions which describe
gravitational wave pulses have amplitudes which only vanish asymptotically.
Many well known soliton wave solutions provide examples of such pulses of
infinite duration. The purpose of the present paper is to consider
gravitational and electromagnetic waves which have distinct wavefronts and to
present some new solutions of this type which have variable polarization.
These may include impulsive or shock waves, or waves with much weaker
discontinuities on their wavefronts.

In general, a solution of Einstein's field equations describes a wave with a
distinct wavefront if there exists a characteristic null hypersurface
(wavefront) which divides the space-time locally into two regions --- an
unperturbed region where the solution represents some given background field,
and a wave region where the solution describes the interaction of some wave
with the background field. The appropriate matching conditions for pure
gravitational waves are those of O'Brien and Synge [1]. The problem of
constructing exact solutions for gravitational waves with distinct wavefronts
can be considered as a special case of the characteristic initial value
problem. For this case, the boundary data on one of two intersecting
characteristic surfaces (i.e. on the wavefront) are determined from the
condition of matching the solution through the wavefront with the given
background field. The boundary data on the other characteristic remain
arbitrary, permitting various wave profiles.

Examples of such waves include the well known class of plane (or {\it pp})
gravitational waves (see \S21 of [2]) which can be used to describe sandwich
waves in a Minkowski background~[3]. These will not be considered further
here. In the context of cylindrically symmetric gravitational waves, the pulse
wave solution of Rosen [4] is also well known (see also [5]). Using standard
coordinates $t$ and $\rho$, this is given by $V=0$ for $t<\rho$ and, for
$t\ge\rho$ by
 $$ V=\int_0^{t-\rho}{F(\sigma)\,\d\sigma\over\sqrt{(t-\sigma)^2-\rho^2}}
\eqno(1.1) $$
 where the function $F(\sigma)$ vanishes outside a finite range $0<\sigma<T$,
and the metric function $V$ is defined below. This solution describes a pulse
gravitational wave with a distinct wavefront $t=\rho$ which can be made
arbitrarily smooth by taking a sufficient number of derivatives of
$F(\sigma)$ to be zero on $\sigma=0$. However, although this defines a
general class of solutions for $V$, the corresponding expressions for all
the metric components cannot be obtained in an explicit form.

In recent years, however, further families of exact solutions have been
obtained for gravitational waves with distinct, possibly curved, wavefronts
which propagate into various backgrounds. For these solutions explicit
expressions for all the metric components have been obtained in closed form. 
Generally, these solutions describe gravitational waves with distinct
wavefronts which propagate either into a Minkowski space or into some other
background of cosmological significance. The wavefronts of these waves may be
plane when propagating into a vacuum Kasner background, or either plane or
cylindrical (or toroidal) when propagating into a Friedmann--Robertson--Walker
universe with a stiff perfect fluid [6--9]. Other families have been obtained
[10--11] which represent gravitational waves with cylindrical, spherical or
toroidal wavefronts propagating into a Minkowski background.

All these space-times admit a two-dimensional abelian group of spacelike
isometries. However, so far, solutions have only been given for the case in
which the metric is diagonal. This case corresponds to that in which the
gravitational waves have constant (linear) polarization. Mathematically, it
corresponds to the case in which the main field equation is the linear
Euler--Poisson--Darboux equation with non-integer coefficients. (In
stationary axisymmetric space-times, this case corresponds to the diagonal
Weyl metrics.) An important consequence of the main equation being linear is
that we can superpose any wave solution on a variety of backgrounds including
a number that are physically significant (e.g. Minkowski in various
coordinates, Kasner, FRW stiff fluid etc.).

The first purpose of the present paper is to clarify the relation between
these solutions and the integral representation of the Rosen pulse (1.1), or
a suitable variation of it. This is achieved in section~3.

The second purpose is to consider the possible generalisation of these
solutions to the nondiagonal case. This problem corresponds physically to the
case in which the gravitational wave has variable polarization, or a
combination of two states of polarization. Mathematically, it corresponds to
the case in which the main field equations can be written as the nonlinear
Ernst equation. It is known that this equation admits various symmetry
transformations which leave it invariant but transform any one solution into
another. Using these transformations it is possible to construct nondiagonal
solutions which involve an arbitrary number of parameters. The problem here
is therefore twofold. It is to construct solutions that correspond to the
physical situation of a wave with a distinct wavefront propagating into a
specific background, and also to express those solutions in closed form.

Since the Ernst equation is nonlinear, it will not generally be possible to
express solutions in the nondiagonal case as a sum of components. Nor will it
be possible to superimpose any particular wave solution on any given
background. Nevertheless, a technique for generating exact solutions of
this type will be presented in section~4 which is based on a pair of linear
equations. Using this, some general classes of wave solutions will be given
explicitly in sections 5 and 6. These represent two classes of gravitational
waves with cylindrical wavefronts propagating into a Minkowski background. An
electromagnetic generalisation of this solution is given in section~7.

\vfil\eject
\noindent{\bf 2. Space-time geometry and field equations}
\medskip
For vacuum space-times with a two-dimensional abelian group of isometries,
with the Killing vectors both spacelike, the line element can be written in
the form
 $$ \d s^2=2e^{-M}\d u\d v-e^{-U}\left(\chi\,\d y^2+\chi^{-1}(\d x-\omega
\d y)^2\right) \eqno(2.1) $$
 in which the functions $U$, $\chi$, $\omega$ and $M$ are functions of the
two null coordinates $u$ and $v$ only. The structure of the vacuum field
equations for this metric is well known. For the function $U$, these
equations imply that 
 $$ e^{-U}=f(u)+g(v), \eqno(2.2) $$ 
 where $f(u)$ and $g(v)$ are arbitrary functions. Two other equations
for the function $M$ can be separated from the main system, namely: 
 $$ \eqalign {
2 U_u M_u =& U^2_u-2 U_{uu}+\chi^{-2}(\chi_u^2+\omega_u^2)\cr
2 U_v M_v =& U^2_v-2 U_{vv}+\chi^{-2}(\chi_v^2+\omega_v^2).} \eqno(2.3) $$ 
 These equations determine $M$ in quadratures provided the main part of the
field equations --- a closed nonlinear system of two equations for metric
functions $\chi$ and $\omega$ --- is solved. These last equations can be
conveniently expressed in terms of a scalar complex potential
 $$ Z\equiv\chi+i\omega \eqno(2.4) $$ 
 in the form which is identical to the well known Ernst equation 
 $$ \hbox{Re}\, Z (2 Z_{uv}-U_u Z_v-U_v Z_u)-2 Z_u Z_v = 0. \eqno(2.5) $$

It may be noted also that, in practice, it is often convenient to treat the
arbitrary functions $f$ and $g$ as null coordinates in place of $u$ and $v$. 
Then, with $e^{-U}=f+g$, the Ernst equation (2.5) takes the form 
 $$ (Z+\bar Z)\left(2Z_{fg}+{1\over{f+g}}(Z_f+Z_g)\right) =4Z_fZ_g. 
\eqno(2.6) $$

For later convenience, we present here also some appropriate expressions for
the space-time curvature of the metric (2.1) in terms of the Newman--Penrose
null tetrad formalism. For a null tetrad defined for (2.1) in a usual form
 $$ \ell^\alpha =e^{M/2}\delta_1^\alpha, \qquad
 n^\alpha =e^{M/2} \delta_0^\alpha, \qquad
 m^\alpha ={1\over\sqrt{2\chi}}e^{U/2}
\big\{(\chi+i\omega)\delta_2^\alpha+i\delta_3^\alpha\big\}  \eqno(2.7) $$
 where $(x^0,x^1,x^2,x^3)\equiv(u,v,x,y)$. 
 The nonzero Newman--Penrose scalars --- the projections on this tetrad of
the selfdual Weyl tensor --- can be expressed as 
 $$ \eqalign { \Psi_0 &={\textstyle{1\over2}}e^U
\big(\partial_v-i\chi^{-1}\partial_v\omega\big)
 \left[{e^{M-U}\over \chi}\partial_v(\chi+i\omega)\right],\cr
 \Psi_2 &={\textstyle{1\over4}}e^M \left[{\partial_u(\chi-i\omega)
\partial_v(\chi+i\omega)\over \chi^2}-U_u U_v\right],\cr
 \Psi_4 &={\textstyle{1\over2}}e^U
\big(\partial_u+i\chi^{-1}\partial_u\omega\big)
\left[{e^{M-U}\over \chi}\partial_u(\chi-i\omega)\right]. \cr} \eqno(2.8) $$ 
 These components and, in particular, their two combinations --- the complex
algebraic invariants of the selfdual Weyl tensor
 $I=2(\Psi_0 \Psi_4+3\Psi_2^2)$, $J=6\Psi_2 (-\Psi_0\Psi_4+\Psi_2^2)$ ---
 describe the properties of the gravitational waves.

When we come to construct solutions for waves with distinct wavefronts, the
appropriate matching conditions must be satisfied across the wavefront.
These are the O'Brien--Synge junction conditions~[1] which, for metrics of
the type (2.1), require the continuity of $U$, $U_u$, $\chi$, $\omega$ and
$M$ across any null hypersurface $u=$const,

\bigskip\bigskip\goodbreak
\noindent{\bf 3. The linear case}
\medskip
Exact solutions for gravitational waves with a distinct wavefront have been
considered extensively in the series of papers [6--11] for the diagonal case,
in which the wave has linear polarization, and for a variety of backgrounds. 
In the linear case, for which $\omega=0$ and the function $Z$ is
real, it is convenient to put  
 $$ \chi=e^{-V}. \eqno(3.1) $$ 
 With this substitution, the Ernst equation (2.6) reduces to the linear
Euler--Poisson--Darboux equation with non-integer coefficients
 $$ (f+g)V_{fg}+{\textstyle{1\over2}}V_f+{\textstyle{1\over2}}V_g=0.
\eqno(3.2) $$

It is well known that the general solution of the characteristic initial
value problem for equation (3.2) admits an explicit integral representation. 
This can be expressed as the sum of two components whose spectral functions
can be related to the corresponding characteristic data by an Abel transform
[12]. Each of these components is equivalent to the form of the Rosen pulse
solution adapted to a more general context. The purpose of this section is to
adapt the corresponding integral representation for the description of
gravitational waves with distinct wavefronts on a given background, and to
clarify the relation between this general form of the solution and the series
of particular exact solutions found  explicitly [6--11].

In these cases, space-times describing gravitational waves with a distinct
wavefront on the null hypersurface $u=0$, on which we will normally have
$f=0$, can be constructed by considering solutions of the form 
 $$ V=V_b+\Theta(u)\tilde V \eqno(3.3) $$
 where $V_b(f,g)$ represents the background space-time, $\Theta(u)$ is the
Heaviside step function which is zero in the background region and $\tilde
V(f,g)$ represents the wave component. In order to satisfy the
O'Brien--Synge junction conditions at the wavefront, it is necessary that
$\tilde V(0,g)=0$.

Assuming that $f\ge0$ and $f+g>0$ behind the wavefront, the general solution
for $\tilde V$ can be written in the form 
 $$ \tilde V(f,g)=\int_0^f
{F(\sigma)\,\d\sigma\over\sqrt{f-\sigma}\sqrt{g+\sigma}}. \eqno(3.4) $$ 
 Putting $f=t-z$ and $g=t+z$, this may be expressed as  
 $$ \tilde V(t,z)=\int_0^{t-z}
{F(\sigma)\,\d\sigma\over\sqrt{t^2-(z+\sigma)^2}}. \eqno(3.5) $$ 
 However, in some geometrical contexts, (such as for cylindrical waves in
which $f=-(t-\rho)\le0$, $g=t+\rho$) $f\le0$, $f+g>0$ behind the wavefront.
In such cases an equivalent integral representations can be given, for which
the equivalent of (3.5) is identical to the Rosen pulse solution~(1.1).

We concentrate here on the case in which $f>0$ in the wave region. It is also
convenient to introduce the new spectral function
$A(\sigma)\equiv F(\sigma)/\sqrt\sigma$. Then, $A(\sigma)$, and hence
$F(\sigma)$, can be determined explicitly in terms of the initial data by
considering this integral on the hypersurface $g=0$ and using the Abel
transform 
 $$ \tilde V(f,0)=\int_0^f {A(\sigma)\,\d\sigma\over\sqrt{f-\sigma}}, 
\qquad \Leftrightarrow \qquad
A(\sigma)= {1\over\pi} \int_0^\sigma {\tilde V_f(f,0)\, \d f
\over\sqrt{\sigma-f}} \eqno(3.6) $$ 
 which was used by Hauser and Ernst [12] in the initial value problem for
colliding plane waves.

In the integral forms of the solution (1.1) and (3.4--5), complete explicit
solutions for the metric function $M$ have not be obtained. However, as shown
in [8] and [11], complete solutions with the required properties can be
obtained by summing over explicit components each of which have the
self-similar form 
 $$ \tilde V_k(f,g)=(f+g)^k H_k\left({\textstyle{g-f\over f+g}}\right) 
\eqno(3.7) $$
 where $k$ is an arbitrary real parameter, and the functions $H_k(\zeta)$
satisfy the linear equation 
 $$ (1-\zeta^2) H_k''+(2k-1)\zeta H_k'-k^2 H_k = 0, \eqno(3.8) $$ 
 with the initial condition $H_k(1)=0$. This satisfies the recursion
relations 
 $$ H_k(\zeta) =\int_1^\zeta H_{k-1}(\zeta')\,d\zeta' \qquad
{\rm so\ that} \qquad H_k'(\zeta)=H_{k-1}(\zeta). \eqno(3.9) $$ 
 Solutions with these properties can be expressed in terms of standard
hypergeometric functions in the form 
 $$ (f+g)^k H_k\left({\textstyle {g-f\over f+g} }\right) 
= c_k\,{f^{1/2+k}\over\sqrt{f+g}}\,
F\left({\textstyle{1\over2}},{\textstyle{1\over2}}\,;
{\textstyle{3\over2}}+k\,;{f\over{f+g}}\right) \eqno(3.10)  $$
 where\footnote{$^1$}
{It may be noted that the expression (3.11) for $c_k$ replaces
incorrect expressions in [10] and [11].}, for integer $k$ 
 $$ c_k= (-1)^k {2^k\Gamma({\textstyle{3\over2}})
\over\Gamma(k+{\textstyle{3\over2}})}. \eqno(3.11) $$ 
 However, when applying (3.9) for arbitrary values of $k$, we only require
the recursion relation $c_{k-1}=-{1\over2}(k+{1\over2})c_k$.

These solutions for each $k$ may be expressed in one of the above integral
forms by first evaluating it on the characteristic $g=0$ on which 
 $$ \tilde V_k(f,0)=f^kH_k(-1). \eqno(3.12) $$ 
 Then, using the Abel transform (3.6), we obtain 
 $$  A(\sigma) ={k\over\pi}H_k(-1) \int_0^\sigma {f^{k-1}\,\d
f\over\sqrt{\sigma-f}}  
={k\over\pi}H_k(-1)\> \sigma^{k-1/2} \int_0^1
{x^{k-1}\,\d x\over\sqrt{1-x}}. \eqno(3.13)   $$ 
 Since the final integral is a constant, it may be concluded that $A(\sigma)$
is a constant multiple of $\sigma^{k-1/2}$. It then follows that the
particular solution (3.7) corresponds to the Rosen form in which the function
$F(\sigma)$ given in (1.1), (3.4), or (3.5) is a constant multiple
of~$\sigma^k$. In addition, it is easy to show directly from (1.1) that
replacing $F(\sigma)$ by some power leads to a self-similar solution. Thus,
substituting $\sigma=t-\rho x$ gives 
 $$ V=\int_0^{t-\rho}{\sigma^k\,\d\sigma\over\sqrt{(t-\sigma)^2-\rho^2}} 
 =\rho^k \int_1^\zeta {(\zeta-x)^k\,\d x\over\sqrt{x^2-1}}. \eqno(3.14) $$ 
 which is an integral representation of $H_k(\zeta)$ for
$\zeta\equiv{t\over\rho}\ge1$.

In the linear case, the general class of solutions which describes waves with
the distinct wavefront $f=0$ can be expressed as a sum of terms of the form
(3.7). This can generally be expressed in the form
 $$ \tilde V= \int_\alpha^\infty \phi(k)\,(f+g)^k
H_k\left({\textstyle{g-f\over f+g}}\right)\, \d k \eqno(3.15) $$
 where $\phi(k)$ is an arbitrary ``spectral amplitude'' function, and the
lower limit $k=\alpha$ is chosen to ensure that there are no curvature
singularities on the wavefront. This particular representation permits the
remaining metric function $M$ to be determined explicitly (see [8] and [11]),
at least for a number of significant backgrounds.

It may be noted that, for a large class of initial data, it is possible to
expand the corresponding $F(\sigma)$ in a Taylor series. In this case it is
only necessary to sum terms of the form (3.7) over integer values of $k>0$.
However, as shown in [9]--[11], in order to include impulses or steps in the
Weyl tensor components it is sometimes necessary to take the minimum value of
$k$ in (3.15) as $\alpha={1\over2}$. These represent impulsive or shock
gravitational waves in some backgrounds. In such situations, it may be
appropriate simply to include other terms with half integer values of $k$.

It has thus been clarified that the above class of solutions (3.3) and (3.15)
can be considered as an alternative and explicit representation of the Rosen
pulse. It is in a form that is applicable to a number of geometrical
backgrounds in which a complete solution for the metric can be obtained
explicitly.

\bigskip\bigskip\goodbreak
\noindent{\bf 4. A simple technique for generating non-diagonal solutions}
\medskip
In the nondiagonal case for which $\omega\ne0$, $Z$ is necessarily complex
and the main equation (2.6) cannot be reduced to a linear form. However, a
number of internal symmetries of the Ernst equation can be used to generate
such solutions.

It is well known that there exists another complex potential ${\cal E}$
(usually called the Ernst potential) which is defined by
 $$ {\cal E}=e^{-U}\chi^{-1}+i\tilde\omega, \qquad \hbox{where} \qquad
\left\{\matrix{\tilde\omega_{u}=e^{-U}\chi^{-2}\omega_u,\hfill\cr
\noalign{\medskip}
\tilde\omega_{v}=-e^{-U}\chi^{-2}\omega_v\hfill\cr} \right. \eqno(4.1) $$
 It is a remarkable property that the function ${\cal E}$ satisfies the same
(Ernst) equation (2.6) as the function $Z$. The identity of the equations
for $Z$ and ${\cal E}$ manifests the existence of a discrete symmetry of the
vacuum Einstein equations for metrics of the form (2.1). This is known as the
Kramer--Neugebauer mapping (or involution): 
 $$ Z\leftrightarrow{\cal E} \qquad {\rm i.e.} \qquad
\chi\leftrightarrow e^{-U}\chi^{-1}, \qquad
\omega\leftrightarrow\tilde\omega. \eqno(4.2) $$
 This mapping is not pure gauge and, in general, it generates from any
given solution another which has a different space-time geometry. A second
application of this symmetry simply leads back to the original solution apart
from an imaginary constant that has no physical significance. According to
this mapping, any solution of the Ernst equation (2.6) can be adopted
either as $Z$ or ${\cal E}$. It can then be shown that, for any vacuum
space-time in which $e^{-U}=f+g$, $Z=\chi+i\omega$ and $M=M_0$, there exists
another solution given by (4.2) in which $e^{-U}=f+g$ is unchanged, and $M$ is
given by 
 $$ M=M_0-{\textstyle{1\over2}}\log(f+g)+\log\chi. \eqno(4.3) $$
 Having different curvatures, these two possibilities correspond to different
physical properties.

Another two very useful groups of symmetries of the vacuum Einstein equations
in this case arise from a continuous group of internal symmetries of the Ernst
equation. These are the SL(2,{\it R}) point transformations of the complex
potentials $Z$ and ${\cal E}$:
 $$ Z\to i{a Z+i b\over c Z+i d} \qquad \hbox{or} \qquad {\cal E}\to
i{a {\cal E}+i b\over c {\cal E}+i d}, \qquad \hbox{both with}
\qquad ad-bc=1. \eqno(4.4) $$ 
 However, there is a very important difference between the corresponding two
SL(2,{\it R}) groups of symmetries. The transformations of the $Z$ potential
are pure gauge transformations for the metric (2.1) which correspond to a
rotation and rescaling of the $x$,$y$ coordinates relative to its Killing
vectors. These leave $M$ invariant for the same $U$ and have no physical
significance. By contrast, the similar SL(2,{\it R}) transformations of the
${\cal E}$ potential, when expressed in terms of the components of the metric
(2.1), can generate changes in the space-time geometry. These latter
transformations are known as Ehlers transformations.

Let us now start with an initial real solution $Z_0=e^{-V}$, where $V$
satisfies (3.2), and take this as an initial solution for ${\cal E}$ namely:
${\cal E}_0=e^{-V}$. Then using the Ehlers transformation, as described
above, a complex solution of the Ernst equation (2.6) can be obtained by
applying a transformation (4.4) for ${\cal E}$ to give 
 $$ {\cal E} = i{{(ae^{-V}+ib)}\over{(ce^{-V}+id)}}, \qquad {\rm with} \qquad
ad-bc=1. \eqno(4.5) $$
 This has real and imaginary parts given by 
 $$ e^{-U}\chi^{-1}={1\over d^2e^V+c^2e^{-V}}, \qquad 
\tilde\omega={bde^V+ace^{-V}\over d^2e^V+c^2e^{-V}}. \eqno(4.6) $$ 
 Substituting this into (4.1) yields the equations for $\omega$ 
 $$ \omega_f=-2cd(f+g)V_f, \qquad \omega_g=2cd(f+g)V_g. \eqno(4.7) $$
 It may be noted that, for nondiagonal solutions, both $c$ and $d$ must be
nonzero and, without loss of generality, it is always possible to put
$c=d=1/\sqrt2$.

The significant and somewhat surprising feature of the equations (4.7) is
that they are linear in $V$, which itself satisfies the linear equation
(3.2). This feature permits arbitrary solutions for $V$ and the corresponding
$\omega$ to be superposed to give a class of solutions of the Ernst equation
in the form 
 $$ Z=(f+g)\cosh V+i\omega \eqno(4.8) $$
 where $V$ satisfies (3.2) and $\omega$ satisfies (4.7) --- all linear
equations. The arbitrary constant which appears in the integration of (4.7)
must be chosen such that the solution can be connected to an appropriate
background solution ahead of the wave. In addition, the solution for $M$
using (4.3) is given explicitly by 
 $$ e^{-M}= \sqrt{f+g} \cosh V \, e^{-M_0} \eqno(4.9) $$ 
 where $M_0$ is the corresponding metric function for the initial diagonal
solution for which $Z_0=e^{-V}$.

\bigskip\bigskip\goodbreak
\noindent{\bf 5. An explicit class of nondiagonal vacuum solutions}
\medskip
 At this point it is convenient to introduce the alternative coordinates
$\tau$ and $\zeta$ defined by
 $$ \tau=f+g, \qquad \zeta={g-f\over f+g}. \eqno(5.1) $$
 Using these variables, equations (4.7) become 
 $$ \eqalign { \omega_\tau &=\tau\zeta V_\tau
-(\zeta^2-1)V_\zeta \cr
 \omega_\zeta &=\tau^2 V_\tau -\tau\zeta V_\zeta. \cr }
\eqno(5.2) $$ 
 Then, putting 
 $$ V=\tau^kH_k(\zeta) \eqno(5.3) $$
 where $H_k(\zeta)$ satisfies (3.8), the equations (5.2) can be integrated to
yield
 $$ \omega ={1\over k+1}\,\tau^{k+1} 
\big[k\zeta H_k+(1-\zeta^2)H_{k-1}\big] \eqno(5.4) $$
 in which the constant of integration has been set to zero to put
$\omega=0$ on the wavefront on which $\zeta=1$. This solution of the
Ernst equation is thus given by
 $$ Z(\tau,\zeta) =\tau\cosh V+i\omega \eqno(5.5) $$
 where $V$ is given by (5.3) and $\omega$ is given by (5.4).

We may now use the property that equations (4.7) for $\omega$, and hence
(5.2), are linear in $V$. It follows that separate solutions for $V$ can be
superposed. We thus obtain a general class of complex solutions of the Ernst
equation in the form (5.5) where 
 $$ V= \int_\alpha^\infty \phi(k)\,\tau^k H_k(\zeta)\, \d k. \eqno(5.6) $$
 and
 $$ \omega =\int_\alpha^\infty \phi(k)\, {1\over k+1}\,\tau^{k+1} 
\big[k\zeta H_k+(1-\zeta^2)H_{k-1}\big]\d k. \eqno(5.7) $$ 
 where $\phi(k)$ is an arbitrary spectral amplitude function, defined over
$[\alpha,\infty)$, subject only to the condition that the above integrals
exist. As explained in~[11], the lower limit $\alpha$ for $k$ must be chosen
such that the space-time is nonsingular on the wavefront. This choice also
determines the behaviour of the Weyl tensor on the wavefront. It will be
shown that, taking $\alpha={1\over2}$, introduces an impulsive wave
component. Gravitational waves with step wavefronts may alternatively occur
if $\alpha={3\over2}$.

It may also be observed that the expression for $\omega$ given by (5.7)
is functionally independent of $V$ as given by (5.6). It follows from this
that it is not possible to introduce a rotation that will diagonalise the
solution globally. It may thus be concluded that these solutions are
genuinely nondiagonal.

We now turn to the question of obtaining an explicit expression for $M$. It
is not necessary to evaluate this by integrating (2.3). Rather, we can use
(4.9) and the expressions given in [11] for the diagonal case to put 
 $$ e^{-M}=|f'g'|\cosh V\,e^{- S} \eqno(5.8) $$
 where 
 $$ \eqalign { S&= -{\textstyle{1\over2}}
 \int_\alpha^\infty \int_\alpha^\infty \phi(k)\phi(k')
{\tau^{k+k'}\over k+k'} \left[kk'H_kH_{k'}
+\left(1-\zeta^2\right)H_{k-1}H_{k'-1}\right] \d k\d k' \cr 
 &= -{\textstyle{1\over2}} \int_{2\alpha}^\infty \d n {\tau^n\over n} 
\int_\alpha^{n-\alpha} \phi(k)\phi(n-k) \left[k(n-k)H_kH_{n-k}
+\left(1-\zeta^2\right)H_{k-1}H_{n-k-1}\right] \d k. \cr } \eqno(5.9) $$ 
 This completes the integration of the field equation for our general class
of solutions.

\bigskip\bigskip\goodbreak
\noindent{\bf 6. Possible backgrounds and the character of the wavefronts}
\medskip
In the above discussion it has been assumed that we are considering a wave
solution with a wavefront given by $f=0$, or $\zeta=1$. For the linear case
considered in~[11], it was convenient to take $V=0$ on the wavefront and to
superpose such a solution on some given background as in (3.3). In the
nonlinear case being considered here, we can similarly superpose a background
onto (5.6). This then introduces additional background terms in (5.7).

For the simplest cases, however, in which the background metric is diagonal,
$\omega$ should be zero for $f\le0$ and the expression for $V$ in the
background must vanish. In this case, no additional terms should appear in
(5.6) and, from (5.5), it can be seen that the solution in the background
region $f\le0$ is given by 
 $$ Z=\tau=f+g, \eqno(6.1) $$ 
 with line element given by
 $$ \d s^2=2|f'g'|\d u\d v-\d x^2-(f+g)^2\d y^2. \eqno(6.2) $$
 It is now necessary to distinguish two cases according to whether the
gradient of $f+g$ is spacelike or timelike.

In the case in which the gradient of $f+g$ is spacelike, it is convenient
to put 
 $$ f=-{\textstyle{1\over2}}(t-\rho), \qquad 
g={\textstyle{1\over2}}(t+\rho). \eqno(6.3) $$ 
 Then, relabelling the other coordinates by putting $x=z$ and
$y=\phi\in[0,2\pi)$, the background is seen to be Minkowski space in
cylindrical coordinates 
 $$ \d s^2 =\d t^2 -\d\rho^2 -\rho^2\d\phi^2 -\d z^2 \eqno(6.4) $$ 
 and the wavefront $f=0$ is the expanding cylinder $\rho=t$. The complete
solution, as constructed above, thus describes a cylindrical gravitational
wave with a distinct wavefront. This can be seen to be an explicit
nondiagonal generalisation of the Rosen pulse in which the complete integral
of all the metric functions has been obtained.

In the alternative case in which the gradient of $f+g$ is timelike, it is
convenient to put 
 $$ f={\textstyle{1\over2}}(t-z), \qquad 
g={\textstyle{1\over2}}(t+z). \eqno(6.5) $$ 
 In this case the background space-time is given by the line element 
 $$ \d s^2=\d t^2-\d z^2-\d x^2-t^2\d y^2. \eqno(6.6) $$
 This can be seen to be one of the particular forms of Minkowski space that
was considered in [11]. Using the results of sections 3 and 4 of [11], the
complete solutions in this case can be seen to describe nondiagonal
generalisations of the solutions given there which represent gravitational
waves propagating into a Minkowski background with half-cylindrical
wavefronts driven by two separating singular null half-planes.

The behaviour of the gravitational wave near the wavefront in each case will
be determined by the component with the lowest value of $k$. We may thus
consider
$V$ to be given by 
 $$ V\approx\phi(\alpha)\tau^\alpha H_\alpha(\zeta)\>\Theta(u) \eqno(6.7) $$
 near the wavefront $u=0$, $f=0$ or $\zeta=1$, and we assume that
$\phi(\alpha)\ne0$. Just behind the wavefront $V$ will be small and using
(5.5), (5.4) and (3.10), to first order, we obtain 
 $$ \eqalignno { Z&\approx\tau +i\omega \cr
 &\approx\tau +i{\phi(\alpha)\over\alpha+1}\tau^{\alpha+1} 
\big[\alpha\zeta H_\alpha+(1-\zeta^2)H_{\alpha-1}\big] \Theta(u) \cr
 &\approx g -i\phi(\alpha)c_\alpha  g^{1/2}f^{\alpha+1/2}\>\Theta(u). &(6.8)
\cr } $$ 
 in which, for integer $\alpha$, $c_\alpha$ is given by (3.11) --- otherwise
$c_\alpha$ may be chosen arbitrarily. Substituting these expressions into
(2.8) we find that, near the wavefront, the non-zero components of the Weyl
tensor are given by 
 $$ \eqalign { \Psi_4 &\approx
{\textstyle{i\over2}{f'\over g'}(\alpha+{1\over2})}
\phi(\alpha)c_\alpha  g^{-1/2}f^{\alpha-1/2}\>\delta(u)
+{\textstyle{i\over2}{f'\over g'}(\alpha^2-{1\over4})} \phi(\alpha)c_\alpha 
g^{-1/2}f^{\alpha-3/2}\>\Theta(u) \cr
 \Psi_2 &\approx {\textstyle{i\over4}(\alpha+{1\over2})}
\phi(\alpha)c_\alpha  g^{-3/2}f^{\alpha-1/2}\>\Theta(u) \cr
 \Psi_0 &\approx {\textstyle{3i\over8}{g'\over f'}} \phi(\alpha)c_\alpha 
g^{-5/2}f^{\alpha+1/2}\>\Theta(u) \cr } \eqno(6.9) $$ 
 It can thus be seen that this class of solutions includes an impulsive
gravitational wave if $\alpha={1\over2}$, the gravitational wave includes a
step (or shock) if $\alpha={3\over2}$, the Weyl tensor is $C^n$ across the
wavefront if $\alpha>n+{3\over2}$.

\bigskip\bigskip\goodbreak
\noindent{\bf 7. Electromagnetic waves with distinct wavefronts}
\medskip
For an electrovac space-time with two Killing vectors, it is known that the
main field equations can be expressed as the Ernst equations 
 $$ \eqalign{ ({\cal R}e\, {\cal E} -H\bar H)\nabla^2{\cal E}
&=(\nabla{\cal E})^2
-2\bar H(\nabla H).(\nabla{\cal E}) \cr
({\cal R}e\,{\cal E} -H\bar H)\nabla^2H &=(\nabla{\cal E}).(\nabla H)
-2\bar H(\nabla H)^2 \cr } \eqno(7.1) $$ 
 where ${\cal E}$ is given by
 $$ {\cal E}=e^{-U}\chi^{-1}+i\tilde\omega+H\bar H. \eqno(7.2) $$
 In this case, $H$ is the electromagnetic potential and ${\cal E}$ is the
generalisation of the vacuum ${\cal E}$-potential which involves the functions
$e^{-U}\chi^{-1}$ and $\tilde\omega$, given by (4.1), rather than the
function $Z$ which contains the explicit metric functions $\chi$ and $\omega$
as in~(2.4).

It is well known that a Bonnor transformation [13] can be used to relate a
class of diagonal electromagnetic solutions to a class of nondiagonal vacuum
solutions. The precise result may be stated in the form that, if $Z_{\rm o}$
is any complex solution of the vacuum Ernst equation (2.5) or (2.6) using
(2.4) with the corresponding $M_{\rm o}$, then a solution (7.2) of the Ernst
equations for an electromagnetic field is given by 
 $$ {\cal E}=Z_{\rm o}\bar Z_{\rm o}, \qquad 
H={\textstyle{1\over2}}(\bar Z_{\rm o}-Z_{\rm o}) \eqno(7.3) $$ 
 and the remaining metric function is given by 
 $$ e^{-M}={4(f+g)^2 \over |f'g'|^3(Z_{\rm o}+\bar Z_{\rm o})^2} 
e^{-4M_{\rm o}}. \eqno(7.4) $$
 The inverse transformation also holds.

We may now apply this to the initial complex solution 
$Z_{\rm o}=(f+g)\cosh V_{\rm o}+i\omega_{\rm o}$ as in (4.8) with the
corresponding $M_{\rm o}$. This then gives an electrovac solution in which 
 $$ {\cal E}=(f+g)^2\cosh^2V_{\rm o}+\omega_{\rm o}^2, \qquad 
H=-i\omega_{\rm o}, \eqno(7.5) $$
 where $V_{\rm o}$ and $\omega_{\rm o}$ are solutions of the linear equations
(3.2) and (4.7). It follows that the new metric functions are given by
 $$ \chi=(f+g)^{-1}{\rm sech}^2V_{\rm o}, \qquad \omega=0, \qquad
e^{-M}=|f'g'|^{-3}{\rm sech}^2V_{\rm o}\,e^{-4M_{\rm o}} \eqno(7.6) $$

Taking $V_{\rm o}$ and $\omega_{\rm o}$ in the forms (5.6) and (5.7), then
$e^{-M}$ takes the form 
 $$ e^{-M}=|f'g'|\cosh^2V_{\rm o}\,e^{-4S} \eqno(7.7) $$ 
 where $S$ is given by (5.9). This solution describes an electromagnetic wave
with wavefront given by $f=0$ propagating into a vacuum background region in
which $\chi=\tau^{-1}=(f+g)^{-1}$. This background is exactly equivalent to
that considered in section~6, and so these solutions describe a combination
of gravitational and electromagnetic waves with cylindrical or
half-cylindrical wavefronts propagating into a Minkowski background.

\bigskip\bigskip\goodbreak
\noindent{\bf 8. Concluding remarks}
\medskip
In the above sections we have effectively given an alternative and explicit
representation of the Rosen pulse solution. This new representation is
significant for a number of reasons. To start with, it enables a complete
integration of the subsidiary field equations, giving an explicit expression
for $M$ in terms of a spectral parameter associated with the pulse. It also
enables the solution to be used in alternative contexts in
addition to cylindrical waves. This then describes a variety of exact
solutions which represent gravitational waves with distinct wavefronts and
which propagate into a number of different backgrounds.

We have also described a method of constructing a particular class of complex
solutions of the Ernst equation which are obtained from a set of linear
equations. This has enabled us to explicitly construct a non-diagonal
generalisation of the Rosen pulse. These describes a family of cylindrical
waves in vacuum or electrovacuum with a distinct wavefront and with two
states of polarization that propagate into a Minkowski background.

\bigskip\bigskip\goodbreak
\noindent {\bf Acknowledgments}
\bigskip 
GAA expresses his thanks to JBG and the Department of Mathematical Sciences
at Loughborough University for hospitality. His visit to Loughborough, where
this work was completed, was supported by a grant from the EPSRC. The work of
GAA was also supported, in part, by the Russian Foundation for Fundamental
Research Grants 96-02-18367 and 96-01-01746 and by the award
INTAS-RFBR95-0435.

\bigskip\bigskip\goodbreak
\noindent {\bf References}
\bigskip

\item{[1]} O'Brien, S. and Synge, J. L. (1952) {\it Commun. Dublin Inst. Adv.
Stud. A}, no 9. 

\item{[2]} Kramer, D., Stephani, H., MacCallum, M. A. H., and Herlt, E.
(1980)  {\it Exact solutions of Einstein's field equations}.
(Cambridge University Press). 

\item{[3]} Bondi, H., Pirani, F. A. E. and Robinson, I. (1959) {\it Proc.
Roy. Soc. A}, {\bf 251}, 519--33. 

\item{[4]} Rosen, N. (1954) {\it Bull. Res. Council Israel.}, {\bf 3},
328--32. 

\item{[5]} Marder, L. (1958) {\it Proc. Roy. Soc. A}, {\bf 244}, 524--37. 

\item{[6]} Griffiths, J. B. (1993) {\it Class. Quantum Grav.} {\bf 10},
975--83.

\item{[7]} Bi\v{c}\'ak, J. and Griffiths, J. B. (1994) {\it Phys. Rev. D},
{\bf 49}, 900--6. 

\item{[8]} Alekseev, G. A. and Griffiths, J. B. (1995) {\it Phys. Rev. D},
{\bf 52}, 4497--502. 

\item{[9]} Bi\v{c}\'ak, J. and Griffiths, J. B. (1996)  {\it Ann. Phys.},
{\bf 252}, 180--210. 

\item{[10]} Alekseev, G. A. and Griffiths, J. B. (1996) {\it Class. Quantum
Grav.}, {\bf 13}, L13--8. 

\item{[11]} Alekseev, G. A. and Griffiths, J. B. (1996) {\it Class. Quantum
Grav.}, {\bf 13}, 2191--209. 

\item{[12]} Hauser, I. and Ernst, F. J. (1989) {\it J. Math. Phys.}, {\bf
30}, 872--887. 

\item{[13]} Bonnor, W. B. (1961) {\it Z. Phys.}, {\bf 161}, 439--44.

\bye